\documentclass[journal]{IEEEtran}
\ifCLASSINFOpdf
\else
\fi

\usepackage{graphicx}
\usepackage{url}
\usepackage{hyperref}
\usepackage{amsmath}
\usepackage{amssymb}
\usepackage{bm}
\usepackage{xcolor}
\usepackage{soul}
\usepackage{afterpage}
\usepackage{subfigure}
\usepackage{caption}

\usepackage{algorithm}
\usepackage{algpseudocode}
\usepackage[colorinlistoftodos]{todonotes}
\usepackage{multirow}
\usepackage{url}

\usepackage{comment}
\usepackage{subfloat}
\usepackage{stfloats}
\usepackage{xcolor}
\usepackage{ulem}
\usepackage{lettrine}
\usepackage{multicol} 
\usepackage{multirow}
\makeatletter
\def\ps@IEEEtitlepagestyle{%
	\def\@oddfoot{\mycopyrightnotice}%
	\def\@evenfoot{}%
}
\def\mycopyrightnotice{%
{\footnotesize }
\gdef\mycopyrightnotice{}
}
\makeatother
\PassOptionsToPackage{draft}{hyperref}
\hypersetup{draft}

\usepackage{fancyhdr}

\fancypagestyle{firststyle}
{
   \fancyhf{}
   
   \fancyhead[C]{\scriptsize{This article has been accepted for publication in IEEE Network. This is the author's post-print version.
   \newline Copyright~\copyright~2021 IEEE. Personal use of this material is permitted. Permission from IEEE must be obtained for all other uses, in any current or future media, including reprinting/republishing this material for advertising or promotional purposes, creating new collective works, for resale or redistribution to servers or lists, or reuse of any copyrighted component of this work in other works.}}
}

\begin{document}
\title{Towards 6G Non-Terrestrial Networks}
\author{Giuseppe Araniti, Antonio Iera, Sara Pizzi, and Federica Rinaldi

\thanks{Giuseppe Araniti, Sara Pizzi, and Federica Rinaldi are with DIIES Dept., University Mediterranea of Reggio Calabria, and CNIT, Italy (e-mail: araniti$|$sara.pizzi$|$federica.rinaldi\}@unirc.it).}
\thanks{Antonio Iera is with DIMES Dept., University of Calabria, and CNIT, Italy (e-mail: antonio.iera@dimes.unical.it).}}

\maketitle

\begin{abstract}
Sixth-Generation (6G) technologies will revolutionize the wireless ecosystem by enabling the delivery of futuristic services through terrestrial and non-terrestrial transmissions.
In this context, the Non-Terrestrial Network (NTN) is growing in importance owing to its capability to deliver services anywhere and anytime and also provide coverage in areas that are unreachable by any conventional Terrestrial Network (TN). The exploitation of the same radio technology could greatly facilitate the integration of NTNs and TNs into a unified wireless system. Since New Radio (NR) is the de facto standard to deliver manifold heterogeneous services in terrestrial wireless systems, 3GPP is investigating new solutions to extend NR to NTNs. In this paper, the constraints that NTN features place on NR procedures are investigated by going thoroughly into 3GPP specifications; strengths and weaknesses of the NR technology in enabling typical 6G services on NR-enabled NTNs are identified; finally, open issues and insights are provided as guidelines to steer future research towards 6G NTNs. 
\end{abstract}


\thispagestyle{firststyle}

\IEEEpeerreviewmaketitle
\section{Introduction}

Fifth-Generation (5G) wireless systems are devised to meet the ever more demanding needs of different emerging service categories with distinct requirements: enhanced Mobile Broadband (eMBB), massive Machine-Type Communications (mMTC), and Ultra-Reliable Low Latency Communications (URLLC). 
However, the time is ripe for a further leap forward towards the provision of novel services, inconceivable up to a few years ago.
Most future Sixth-Generation (6G) applications will result from the merging of the 5G usage categories, as described, for example, in the ITU NET-2030 focus group White Papers.
Specifically, 6G will mainly support: \textit{Holographic Type Communications (HTC)} to project holographic subjects inside a remote place; \textit{Multi-Sense Networks} to create fully immersive experiences by involving all the senses; \textit{Time Engineered Applications} where {machines and sensors can react to unpredictable and unprogrammed events and autonomously and quickly solve encountered problems without human intervention}; and \textit{Critical Infrastructure} to ensure safety whenever an emergency occurs.

Recently, the interest in Non-Terrestrial Networks (NTNs) \cite{ieee_access} by both academia and industry has tremendously increased, and commercial solutions began to appear. 
For instance, SpaceX Starlink delivers satellite broadband services in unreachable areas with low latency (i.e., below 30 ms) and high individual data rates (i.e., above 100 Mbps), and the coverage  is significantly extended with an average of two Starlink launches per month.

NTN integration within the depicted 6G wireless ecosystem is essential to ensure service availability, continuity, ubiquity, and scalability. 
However, NTN distinguishing features (e.g., orbit type, altitude, footprint size) strongly depend on the NTN platform type, spanning from Geostationary/Medium/Low Earth Orbit (GEO, MEO, LEO) satellites to Unmanned Aircraft Systems (UAS), including High Altitude Platform Systems (HAPS).
Since the TN suffers from limitations in terms of deployment and coverage, the NTN may be a complement to achieve global connectivity, e.g., through nano-satellite constellations. Undoubtedly, the deployment of a unified wireless system including both TNs and NTNs could be greatly facilitated by the exploitation of the same radio technology in both Radio Access Networks (RANs). 

{TN-NTN integration has been the subject of recent survey papers. Some works focus, on a broad spectrum, on architectures, use cases, enabling technologies of NTN in 5G/6G, and highlight the possible role played by satellites in future terrestrial-satellite platforms. Others instead refer to 3GPP specifications and focus on user plane and control plane related problems and solutions. Differently, the objective of our paper is to trace the needs that will guide the transition from NTN platforms in 5G/beyond-5G based on New Radio (NR) \cite{wir_com} technology towards NTN platforms integrated into 6G systems. Obviously, an alternative ``clean slate" approach to the design of 6G NTNs can be followed. Our paper is supporting the idea that, at least during their early stages, 6G NTNs may be evolved from 5G NTNs. Therefore, we analyze the main requirements of future 6G use-cases, investigate the impact of NTN constraints on NR procedures, and provide an exhaustive vision of the main aspects of current NR that must be addressed to overcome the limitations of NTNs in supporting future services. Last, insights are given on the role played in future NTN systems by other 6G key enabling technologies, which will initially add to NR and later, maybe, will replace it.}

In details, Section \ref{sec:2} describes 6G usage scenarios and applications. 
Section \ref{sec:3} and \ref{sec:4} summarize, respectively, the NTN architecture and the NR technology.
Section \ref{sec:5} discusses the main reasons to adopt NR for NTNs.
Section \ref{sec:6} investigates the impact of NTN features and constraints on NR procedures by going thoroughly into 3GPP specifications. 
Section \ref{sec:7} outlines the key enablers that will foster the development of future integrated space-air-ground networks. 
Conclusions are drawn in Section \ref{sec:8}. 

\begin{figure}[h]
\centering
{\includegraphics[scale=0.34]{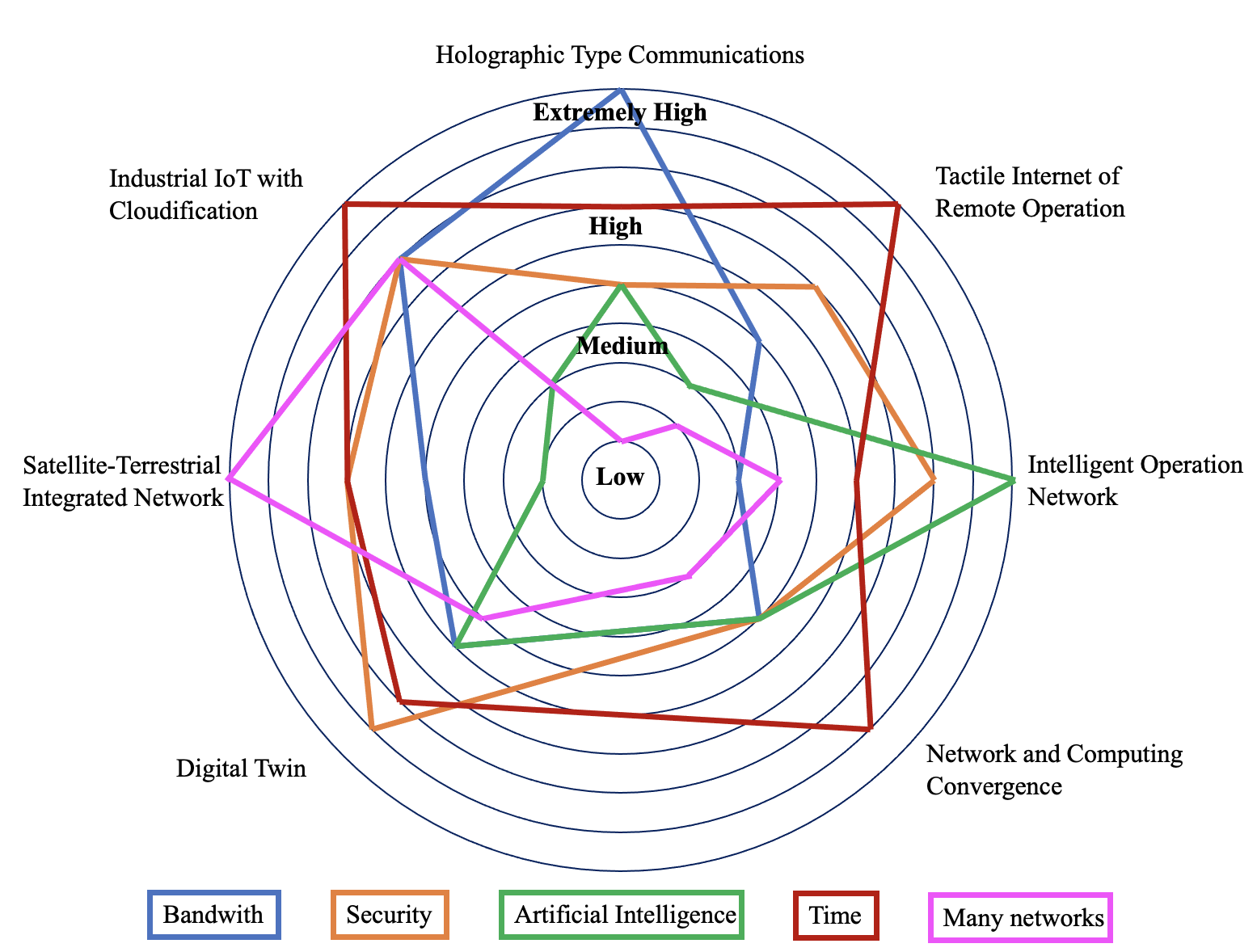}}
\caption{{Importance levels of requirements for 6G use cases.}}
\label{fig:6gusecases}
\end{figure}

\section{6G Use Cases and Usage Scenarios}
\label{sec:2}

The 6G era will see the advent of revolutionary applications. Fig. \ref{fig:6gusecases}, taken from the ITU-T technical report on ``Representative Use Cases and Key Network Requirements for Network 2030", clearly illustrates the importance levels of key network requirements for the most representative 6G use cases:
\begin{itemize}
    \item \textbf{\textit{HTC}} realizes fully immersive experiences through the three-dimensional (3D) projection of objects or people.
    \item \textbf{\textit{Tactile Internet of Remote Operation}} enables real-time control in Industry 4.0 (i.e., remote industrial management) and telemedicine (i.e., remote robotic surgery). 
    \item \textbf{\textit{Intelligent Operation Network}} exploits Artificial Intelligence (AI) capabilities that simulate the human brain to fix network malfunctions and faults.
    \item \textbf{\textit{Network and Computing Convergence}} leads to the interconnection of multiple distributed network edge sites. 
    \item \textbf{\textit{Digital Twin}} represents real-time objects, processes, and people with a digital counterpart.
    \item \textbf{\textit{Satellite-Terrestrial Integrated Network}} provides ubiquitous Internet coverage, edge caching and computing services, and more network access paths. 
    \item \textbf{\textit{Industrial Internet of Things (IIoT) with Cloudification}} offers secure, reliable, and real-time intra-/inter-factory connectivity without human intervention.
\end{itemize}

{Terms of comparison are: \textit{Bandwidth} needed to deliver specific applications; \textit{Security}, including data integrity, privacy protection, trustworthiness, resilience, lawful interception, and traceability; \textit{AI} capability to autonomously analyze the network status and enable intelligent applications; \textit{Time}, expressing the relevance of reduced latency and real-time geolocation mechanisms when considering mobility; \textit{Many networks}, accounting for the seamless coexistence of many heterogeneous networks for global connectivity.}

6G applications can be classified in \textit{ubiquitous Mobile Ultra-Broadband (uMUB)}, \textit{ultra-High Data Density (uHDD)}, and \textit{ultra-High Speed and Low-Latency Communications (uHSLLC)} \cite{VTM_2019}. Since 6G usage scenarios are characterized by dissimilar requirements and NTNs include a wide range of spaceborne/airborne categories, it is of great importance to evaluate which kind of NTN platform best suits each service macro-category.

\textbf{\textit{uMUB}} requires extremely high peak data rates (i.e., $\geq$ 1 Tbps in 6G vs 0.02 Tbps in 5G) and experienced data rates (i.e., 1 Gbps in 6G vs 0.1 Gbps in 5G), very high traffic capacity (i.e., 1 Gbps/m$^2$ in 6G vs 0.01 Gbps/m$^2$ in 5G), seamless coverage, and mobility (i.e., $\geq$ 1000 km/h in 6G vs 500 km/h in 5G). Owing to wide-area coverage, the NTN provides broadband connectivity over areas un-served or under-served by TNs by supporting very-high-speed mobility with limited handover mechanisms. 
In particular, NTN systems based on High Throughput Satellite (HTS) and Very-HTS (VHTS) can improve throughput, spectral efficiency, and system capacity. GEO HTS can satisfy the above-mentioned high data rate requirements but at the expense of a longer propagation delay, while LEO HTS may offer the best trade-off between data rates and latency in the case of uMUB applications. However, LEO satellites movements around Earth, call for managing frequent handovers and adopting new solutions to compensate for the Doppler effect.

{\textbf{\textit{uHDD}} requires high reliability and ultra-high density of IIoT devices (i.e., 10$^7$ devices/km$^2$ in 6G vs 10$^6$ devices/km$^2$ in 5G). Considering the wider satellite footprint size compared to terrestrial cell size and the emerging need for delivering IoT devices also in rural areas where the TN coverage is partially or completely missing, leveraging NTN constellations to meet demands for uHDD services seems a highly suitable solution. 
Again, whenever applications, such as factory automation and machine control, are involved that require real-time, secure, and reliable connectivity among IIoT devices, the most appropriate NTN technology is represented by LEO satellites that may overcome the issue of long propagation delay generated by GEO satellites and, at the same time, may serve dense populations of devices also in remote areas.}
    
\textbf{\textit{uHSLLC}} require ultra-high data rates and very low latency (i.e., 0.01-0.1 ms in {ideal} 6G vs 1 ms in 5G). uHSLLC also give high importance to mobility.
{Hierarchical NTN architectures with airborne vehicles (i.e., Unmanned Aerial Vehicle - UAV, and HAPS) deployed on several levels may be an interesting solution to meet the different constraints on the required latency {\cite{ST_2021}}.}
Conversely, the very high-speed mobility has an impact on localizing devices, planning the radio frequencies, and coordinating multiple airborne-based NTN platforms over a certain area of interest. Moreover, new techniques should be implemented to enhance reliability. These issues make the provision of uHSLLC via NTN a challenging task.

\section{NTN Architecture} 
\label{sec:3}

{In 3GPP TR 38.821, the NTN access encompasses \textit{NTN terminals}, \textit{NTN platforms}, and \textit{NTN Gateways (GWs)}.}

{The NTN terminals are 3GPP User Equipments (UEs) or non-3GPP satellite terminals with or without Global Navigation Satellite System (GNSS) support. They operate in Ka-band or S-band and are connected to the NTN platform through the \textit{service link}. NTN platforms generate several fixed or moving beams of elliptic shape over a given area (i.e., footprint) and embark a transparent or a regenerative payload. In the latter case, NTN platforms may be interconnected through \textit{inter-satellite links (ISLs)}. The NTN GWs are connected to NTN platforms through \textit{feeder links}.}

{3GPP has defined the NR-enabled NTN architecture that minimizes the need for new interfaces and protocols in next-generation RAN (NG-RAN) to support NTNs.}

{Specifically, in \textit{transparent NTN}, there is no need to modify the NG-RAN architecture; indeed, the NR-Uu radio interface of the feeder link is repeated in the service link, and NR-Uu timers are extended to cope with the long propagation delay of the feeder and service links. In \textit{regenerative NTN}, the satellite implements functions of a gNB, thus increasing the delay over the feeder link and needing the extension of NG Application Protocol timers. 
The NG interface of the feeder link is over Satellite Radio Interface (SRI). Finally, in \textit{regenerative NTN with functional split}, specified by 3GPP TS 38.401, the distributed unit (DU) of the gNB is on board the satellite, while the central unit (CU) is on the ground. The logical interface F1 over the SRI is on the feeder link connecting the DU to the same CU. This implies a long delay from the DU to the CU, which  requires the F1 timer extension.}

\begin{figure}[h]
\centering
{\includegraphics[scale=0.38]{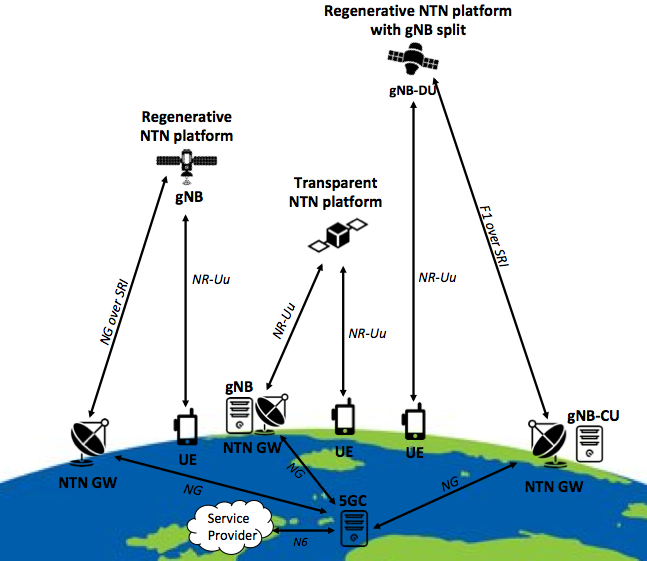}}
\caption{NR-enabled NTN access.}
\label{fig:architettura}
\end{figure}

{Figure \ref{fig:architettura} illustrates the NR-enabled NTN access scenario with the interfaces allowing the NR signal transmission from NTN terminals to 5G Core Network (5GC) and vice versa.}

However, 3GPP Release 17 focuses on NR-enabled NTNs, which are based on LEO and GEO satellites, support HAPS and air-to-ground scenarios, and are assumed to carry a transparent payload. In future 3GPP releases, further studies are also expected on other architecture options.

\section{NR in a Nutshell}
\label{sec:4}

\subsubsection*{\textbf{Physical (PHY) Layer}} NR novelties related to the PHY layer are classified in the following categories.

\subsubsection{Waveform and Scalable Numerology} 
NR introduces the scalable Orthogonal Frequency Division Multiplexing (OFDM) numerology that specifies {Subcarrier Spacing (}SCS{)} from 15 to 480 kHz, Transmission Time Interval (TTI) from 1 ms to 31.25 $\mu$s, Cyclic Prefix (CP), and number of slots. 
As specified in the 3GPP TS 38.214, to cope with the consequent higher device energy consumption, each device monitors only one active \textit{Bandwidth Part} (BWP), which is a fixed band correlating the numerology and the scheduling mechanism. 

\subsubsection{Frame Structure}
The NR frame structure supports Frequency Division Duplex (FDD) and Time Division Duplex (TDD) during transmissions in the paired spectrum and unpaired spectrum, respectively. 
An NR subframe is formed by adjacent slots, each comprising 7 or 14 OFDM symbols for SCS $\le 60$ kHz and 2, 4, 7, or 14 OFDM symbols for SCS $\le 120$ kHz, as described in 3GPP TS 38.213.
NR considers mini-slots to be the smallest scheduling units of 2, 4, or 7 symbols and important for uHSLLC.

\subsubsection{Multi-Antenna Transmission and Beamforming}
The adoption of a large number of antenna elements, multi-user Multiple Input Multiple Output (MU-MIMO), and beamforming, increase the transmission/reception capability and overcome the signal attenuation 
at high frequencies. 
The beam management establishes the correspondence between the directions of transmitter and receiver beams by identifying the most suitable beam pair.

\subsubsection*{\textbf{Medium Access Control (MAC) Layer}} 
The procedure for creating MAC PDUs, known as Logical Channel Prioritization (LCP), is enhanced due to scalable numerology and TTIs.
The LCP handles the mapping of a logical channel onto certain numerology/TTI and includes uHSLLC traffic prioritization.
Beyond the improved PDU format, NR MAC adopts new scheduling request (SR) functionality. The SR is utilized to request Uplink Shared Channel (UL-SCH) resources to establish a data transmission. 
The NR MAC entity is configured such that a UE may support zero, one, or more SR configurations.

\subsubsection*{\textbf{Radio Link Control (RLC) Layer}} 
The NR RLC does not order the Service Data Units (SDUs) in sequence to reduce the overall latency; hence, packets can be forwarded towards the upper layers immediately, without waiting for re-transmissions of previously missing packets. To meet the NR latency requirements, concatenation is not supported.

\subsubsection*{\textbf{Packet Data Convergence Protocol (PDCP) Layer}} 
The re-ordering functionality is moved from RLC to PDCP{, which} is also in charge of ciphering, deciphering, and integrity protection, as well as duplication.
Accordingly, user packets are re-transmitted to the gNB several times such that at least one copy is received correctly. If more than a single copy of the same PDU is received, the PDCP removes any duplicates. PDCP duplication is also used (i) in case of transmission via multiple cells to meet high-reliability requirements and (ii) in dual-connectivity scenarios.

\subsubsection*{\textbf{Service Data Adaptation Protocol (SDAP) Layer}} 
The new SDAP manages QoS when connected to the 5GC. Each PDU session 
consists of QoS flows and data radio bearers (DRBs) and SDAP provides the mapping of a QoS flow from the 5GC and a DRB while marking the QoS Flow Identifier.

\subsubsection*{\textbf{Radio Resource Control (RRC) Layer}} 
The RRC has solely control-plane functionalities. It operates between a device and its gNB. It handles the paging procedure; establishes, maintains, and releases the RRC connection between a UE and NG-RAN by including security, mobility, and QoS management functions. The RRC also helps detect and recover from radio link failures. 
Each NR device supports three RRC states: NR introduces INACTIVE as an intermediate state between IDLE and CONNECTED. 

\section{NR to Support 6G applications over NTNs:\\ Motivations and Concerns}
\label{sec:5}

{In this section, we discuss advantages, disadvantages, and potential solutions of enabling NR over NTNs to support 6G applications, as summarized in Table \ref{tab:ref_NR-6Gapp}.}

\begin{table*}[t!]
\small
\caption{{{Summary of advantages, disadvantages, and solutions to enable NR over NTNs to support 6G applications.}}}
\label{tab:ref_NR-6Gapp}
\centering
{
\begin{tabular}{|p{1cm}|p{4.5cm}|p{4.5cm}|p{4.5cm}|}
\hline
\multicolumn{1}{|c|}{\multirow{3}{*}{\textbf{NR novelties}}} & \multicolumn{1}{c|}{\multirow{3}{*}{\textbf{Advantages}}} & \multicolumn{1}{c|}{\multirow{3}{*}{\textbf{Disadvantages}}} & \multicolumn{1}{c|}{\multirow{3}{*}{\textbf{NTN Solutions}}}\\
& & &\\
& & &\\
\hline
\multicolumn{1}{|c|}{\multirow{4}{*}{\textit{mmWave}}} & 
$\cdot$ Extreme capacity\newline
$\cdot$ Elevated throughput\newline
$\cdot$ Low latency
& 
$\cdot$ Reduced coverage\newline
$\cdot$ High signal attenuation
& 
Select operating frequencies according to application requirements; mobility of NTN terminals and platforms; NTN platform altitude; propagation channel.
\\
\hline
\multicolumn{1}{|c|}{\multirow{3}{*}{\textit{Scalable numerology}}} & 
$\cdot$ Network scalability\newline
$\cdot$ Network flexibility\newline
$\cdot$ Support any service type
& 
Long propagation delay for the adoption of high numerology when providing uHSLLC.
& 
Consider low-altitude NTN platforms to not add significant delay to the delivery of uHSLLC service.\\
\hline
\multicolumn{1}{|c|}{\multirow{5}{*}{\textit{Design}}} & 
$\cdot$ Increased energy efficiency\newline
$\cdot$ Reduce always-on transmissions\newline
$\cdot$ Reduced energy consumption\newline
$\cdot$ Boosted battery life\newline
$\cdot$ Fast and reliable transmissions
& 
Computational complexity in the beam-pair selection, establishment, and maintenance.
& 
$\cdot$ Enable the inactive state of devices\newline
$\cdot$ Adopt multiple antennas and beamforming scheme\\
\hline
\multicolumn{1}{|c|}{\multirow{4}{*}{\textit{Supplementary Uplink}}} &
$\cdot$ Extended uplink coverage\newline
$\cdot$ Increased uplink data rates
& 
-
&
Consider supplementary uplink for low-altitude NTNs to extend uplink coverage during the reception of uHDD and uHSLLC messages.\\
\hline
\end{tabular}
}
\end{table*}

\subsubsection*{\textbf{Radio spectrum}} 
NR operates in both lower bands (i.e., below 7 GHz in FR1) and mmWave (i.e., between 24.25 GHz and 52.6 GHz in FR2). Operating at mmWave frequencies guarantees extreme capacity, throughputs of multiple Gbps, and low latency. 
{Conversely, the signal is subject to a greater attenuation and mobility management is more challenging.}
{This phenomenon, accentuated in tough propagation conditions, implies a reduction of the coverage area size and a consequent increase of handover requests. This issue needs to be investigated to take advantage of the benefits introduced by the use of extremely high frequencies when delivering uMUB, uHDD, and uHSLLC services.} The selection of operating frequencies is challenging due to further factors, such as application requirements, mobility of NTN terminals/platforms, altitude of NTN platforms, and propagation channel.

\subsubsection*{\textbf{Scalable numerology}} 
The scalable numerology enables network flexibility and scalability in supporting any kind of service application through a proper selection of SCS and TTI parameters. Low numerologies, characterized by narrow SCS and long TTI, may be used for uMUB service delivery. High numerologies with wider SCS and shorter TTI may be exploited for uHDD, where each sensor transmits a few data, and for uHSLLC that need very low latency transmissions. 
However, in NTN the adoption of high numerology to support uHSLLC may be not enough, due to the long propagation delay. Very low-altitude NTN platforms may be a solution not to add significant delay to the uHSLLC transmission. 

\subsubsection*{\textbf{Design}}
NR is designed \textit{ultra-lean} to increase energy efficiency by reducing always-on transmissions. This feature is particularly important for uHDD where sensors need low energy consumption for long battery life. Furthermore, the introduction of the inactive state of devices is a key enabler for applications requiring low latency and minimum battery consumption. 
Moreover, NR is \textit{beam-centric}, which means that beamforming and multi-antenna schemes are extended from data transmission to control-plane procedures and initial access. 
Beamforming may allow fast and reliable uHSLLC. 
Finally, NR ensures forward compatibility to future use cases and technologies that may be supported also by NTNs. 

\subsubsection*{\textbf{Supplementary Uplink (SUL)}}
NR SUL associates a conventional downlink/uplink carrier with a supplementary uplink carrier operating at lower frequencies. The objective of SUL is to extend uplink coverage and increase uplink data rates in the case of limited power owing to reduced path loss in low-frequency bands. This may be useful to extend the coverage of low-altitude NTN platforms during the reception of uHDD and uHSLLC messages. 

\section{Impact of NTN on NR Procedures}
\label{sec:6}

\begin{table*}[h!]
\small
\caption{{Enhancements to user-plane and control-plane NR protocols to support NTNs.}}\label{tab:ref_table2}
\centering
{
\begin{tabular}{|p{1cm}|p{16cm}|}
\hline
\multicolumn{1}{|c|}{\multirow{3}{*}{\textbf{Layer}}} & \multicolumn{1}{|c|}{\multirow{3}{*}{\textbf{Modifications to NR protocols}}} 
\\
&\\
&\\
\hline
\multicolumn{1}{|c|}{\multirow{9}{*}{\textbf{PHY}}} & 
$\cdot$ Introduce an offset to timers in DL/UL frame timing (i.e., for transmission on Physical Uplink Shared CHannel (PUSCH), HARQ-ACK feedback, Channel State Information (CSI), aperiodic Sounding Reference Signal (SRS), MAC Control Element (CE) action, and CSI reference resource);

$\cdot$ Investigate beam management, BWP operations, signaling of polarization mode, and impacts of feeder link switch on PHY layer procedures;

$\cdot$ Enhance Physical Random Access CHannel (PRACH) formats and preamble sequences, e.g. adopting repetitions or larger sub-carrier spacing;

$\cdot$ Modify the number of HARQ processes that consider HARQ feedback, HARQ buffer size, RLC feedback, and RLC ARQ buffer size.
\\
\hline
\multicolumn{1}{|c|}{\multirow{8}{*}{\textbf{MAC}}} & 
$\cdot$ Extend preamble receiving window and random access response;

$\cdot$ Introduce additional offsets for delaying the start of the random access response window and the contention resolution timer;

$\cdot$ Consider other random access enhancements to address mobility issues;

$\cdot$ Add an offset for Discontinuous Reception (DRX) to save NTN terminal battery;

$\cdot$ Consider a longer DRX inactivity timer and a short DRX cycle;

$\cdot$ Extend the range of the prohibit timer, which indicates when the NTN terminal can use a scheduling request;

$\cdot$ Enhance transmission scheduling.
\\
\hline
\multicolumn{1}{|c|}{\multirow{3}{*}{\textbf{RLC}}} & 
$\cdot$ Extend the timer for detecting the Protocol Data Unit (PDU) reception failure;

$\cdot$ Extend the RLC Sequence Number (SN) length;

$\cdot$ Reduce the delay to perform RLC retransmissions. 
\\
\hline
\multicolumn{1}{|c|}{\multirow{3}{*}{\textbf{PDCP}}} & 
$\cdot$ Modify the timer of PDCP Service Data Unit (SDU) discard and the timer of loss detection of PDCP Data PDU; 

$\cdot$ Extend the PDCP SN length;

$\cdot$ Reduce the delay to perform PDCP retransmissions.
\\
\hline
\multicolumn{1}{|c|}{\multirow{1}{*}{\textbf{SDAP}}} & 
--
\\
\hline
\multicolumn{1}{|c|}{\multirow{4}{*}{\textbf{RRC}}} & 
$\cdot$ Enhance connected mode mobility procedures to address the large propagation delay (e.g., by considering aspects related to the satellite movement in case of NGSO satellite);

$\cdot$ Consider solutions for identifying NTN moving footprints and generating steerable beams, whose footprints are fixed on the ground.
\\
\hline
\end{tabular}
}
\end{table*}

{In this section, we investigate how to adapt NR procedures to cope with NTN typical features/constraints. For the reader's convenience,  a summary of the needed adjustments to NR is also given in Table \ref{tab:ref_table2}.}

\subsubsection*{\textbf{Propagation Delay}}
The long propagation delay implies changes to the following mechanisms:
\begin{itemize}
\item \textit{Timing Advance (TA):}
dynamic update and alignment of UE individual TAs to ensure the synchronization of all UL transmissions at the gNB and to compensate for the large TA offset in DL and UL frame timing.
\item \textit{Hybrid Automatic Repeat Request (HARQ):}
adaptations could include (i) the increase of periodicity and number of symbols retransmissions to satisfy requirements of high reliability in the case of low/moderate propagation delays and (ii) the decrease/disabling of HARQ processes in the case of long propagation delays. 
\item \textit{Adaptive Modulation and Coding (AMC):}
selects the most robust modulation to compensate for the large response time due to the long propagation delay. This leads to a reduction in spectral efficiency.
\item \textit{MAC and RLC procedures:} new methods for (i) sending data and access signaling during the Initial Access procedure, (ii) implementing flexible or extended receive window size, flexible ARQ/HARQ cross coordination, radio resource allocation, ACK free scheme or a latency adaptive HARQ-ACK feedback to cope with a long delay channel during the Data Transfer procedure. 
\end{itemize}

\subsubsection*{\textbf{Propagation Channel}}
The propagation channel of spaceborne platforms (i.e., GEO, MEO, LEO) is modeled according to the Ricean distribution with a strong LoS signal component and slow fading is possible in temporary signal masking (e.g., under trees and bridges), whereas the signal belonging to airborne platforms (i.e., UAS and HAPS) is given by multipath components with frequent and fast fading. 
To compensate for the Doppler effect, it may be necessary to adapt the synchronization configuration of both UE and gNB receivers, the reference signals in the physical signals, and the preamble sequence and aggregation. The adaptation of the cyclic prefix, wherein multiple multi-path components are recombined, may compensate for the delay spread, whereas the SCS extension to greater values may adapt the OFDM signal in situations of larger Doppler effect. 
{In LEO-based NTNs, obtaining an accurate instantaneous CSI (iCSI) of NTN terminals is practically unfeasible due to long propagation delay and terminal/satellite mobility, whereas it is easier to gather the statistical CSI (sCSI) that varies slower than iCSI. Massive MIMO schemes exploiting sCSI with Full Frequency Reuse (FFR) overcome the issue of iCSI collection, reduce the computational overhead, and increase the experienced data rate in LEO-based NTNs \cite{You}.}

\subsubsection*{\textbf{Frequency Plan and Channel Bandwidth}} NTN terminals operate in Ka-band and S-band frequencies. 
The review of the carrier numbering and the configuration of the 5G radio inference mode is needed to support (i) the operational spectrum, (ii) the pairing between UL/DL bands with specific band separation, and (iii) the FDD access scheme in Ka-band deployment scenarios. Other adaptations may be needed for the MAC and network layer signaling.
Besides the extension of the bandwidth up to 800 MHz, at Ka-bands an alternative method to provide a similar throughput is carrier aggregation, which offers  flexibility in allocating carriers among cells while satisfying frequency reuse constraints.

\subsubsection*{\textbf{{Link Budget}}} 
{The Peak to Average Power Ratio (PAPR) is an important measure for link budget since it determines the vulnerability of the transmitted signal to non-linear distortion due to constellation warping and clustering:  the higher the PAPR, the greater the non-linear distortion can be. To reduce this effect,} (i) the power backoff of the amplifier shall be increased close to the saturation point even if this leads to a reduction in terms of the amplifier efficiency, (ii) signal processing techniques may be performed, and (iii) the {Modulation and Coding Scheme (}MCS{)} may be extended. 
Besides offering greater robustness against distortion, extending the  MCS of NR for low SNR is a way to increase the signal availability, especially for cell-edge NTN terminals in case of slow and deep fading conditions, and to meet the reliability requirements of critical communications.
{Furthermore, a proper selection of the NR numerology overcomes the issues related to the effects of power-limited link budget and high Doppler in LEO-based NTNs and integrated terrestrial/NTN systems \cite{Jayaprakash}.}

\subsubsection*{\textbf{Cell Pattern Generation}}
The NTN platform footprint size has an impact on initial access procedures. 
When NTN terminals are unaware of their positions, large footprint size entails {} long differential delay between two chosen points thus generating the near-far effect, which can be limited by extending the acquisition window.
On the contrary, when NTN terminal positions are known, the differential delay {as well as the Doppler effect caused by the NTN platform mobility} can be compensated by the network {through the implementation of techniques based on the GNSS when the UE is equipped with the GNSS receiver.}
Even when the NTN differential delay is lower or equal to the cyclic prefix duration specified for the NR Physical RACH (PRACH), the residual differential delay may be mitigated; otherwise, further studies are needed to find alternative solutions for the NR RACH procedure \cite{Zhen_6GLEO}. 

\subsubsection*{\textbf{NTN Platform Mobility}}
The NTN platform mobility leads to specific cell patterns (i.e., size and position), while the NTN footprint may cross boundaries between countries. This impacts cell identification methods, tracking and location area design, roaming and billing procedures, and location-based service delivery. 
Some adaptations to the inter-gNB protocols of specific NTN topologies may be required for handling the propagation delay, cell pattern, and cell mobility due to the NGSO satellite motion.
Moreover, moving cells entail a new open issue on the paging capacity due to the unacceptable overhead of Registration Area Update requests during network-initiated calls. On one hand, designing a fixed tracking area where NGSO satellites steer the beams may be a solution to cover a certain area for as long as possible. On the other hand, fixed tracking areas are also possible for NGSO satellites with moving beams by changing the tracking area code when arriving in the next Earth-fixed tracking area. Furthermore,
{c}onsidering extended NR SCS values and large channel bandwidth (up to 800 MHz) may be useful to mitigate the Doppler shift due to the NTN platform mobility, especially for Ka-band. 
Table \ref{tab:SCS-bands} shows NR SCS values supported for NTN operational frequencies.

\begin{table}[h]
\caption{NR SCS values supported for NTN operational frequencies.}\label{tab:SCS-bands}
\centering
{
\begin{tabular}{|l|c|c|c|c|c|}
\hline
\multicolumn{1}{|c|}{\multirow{4}*{\textbf{Operational Frequency}}}&  \multicolumn{4}{|c|}{\multirow{2}*{\textbf{NR SCS values}}}\\
 & \multicolumn{4}{|c|}{}\\
\cline{2-5}
 & \multicolumn{1}{|c|}{\multirow{2}*{\textbf{15 kHz}}} & \multicolumn{1}{|c|}{\multirow{2}*{\textbf{30 kHz}}}& \multicolumn{1}{|c|}{\multirow{2}*{\textbf{60 kHz}}}
 & \multicolumn{1}{|c|}{\multirow{2}*{\textbf{120 kHz}}}\\
 & & & &\\
\hline
S-band @2GHz & \checkmark & \checkmark & -- & -- \\
\hline
Ka-band @20GHz (DL) & -- & -- &\checkmark & \checkmark \\
\hline
Ka-band @30GHz (UL) & -- & -- & \checkmark & \checkmark \\
\hline
\end{tabular}
}
\end{table}

\subsubsection*{\textbf{NTN Terminal Mobility}}
Doppler shift generated by NGSO satellites is due to the NTN platform movement, whereas{,} for GEO and HAPS, the main contribution is due to the NTN terminal movement. 
The higher Doppler shift caused by very high-speed mobility could be mitigated by considering wider SCS values.
Furthermore, the NTN terminal mobility leads to physical layer{-}related issues when NTN terminals are moving at very high speeds (i.e., 1000 km/h), such as high power control loop response that can be reduced by: (i) considering mini-slots for the transmission and extending SCS values, and (ii) revisiting the mapping and scheduling of the power control command on physical radio resources.
Finally, GEO satellite NTNs are characterized by a long propagation delay that impacts the random access procedure, specifically, the random preamble delivery \cite{Guidotti2020}. 

\section{Future Research Directions}
\label{sec:7}

New 6G services will be oriented to fully sensory, completely automated, and real-time experiences \cite{IEEE_NETWORK} enabled by a dynamic, reconfigurable{,} and flexible 6G design. 

We envision that soon NTNs will not merely support the terrestrial network, rather it will be a key component of future integrated Telecommunication platforms and several services will be delivered over the space-air connections of the NTN segment.

To offer typical 6G data traffic and to guarantee service availability, ubiquity, and scalability, future systems based on heterogeneous RANs (NTN/TN) shall leverage the 6G key enablers described below.

\textbf{\textit{Terahertz (THz) communications}} achieve the multi-gigabits-per-second capacity required by 6G applications, but suffer from high path loss and molecular absorption and are subject to blockage.
Despite the notable advancements in manufacturing THz devices, there is still a long way to go for the development of end-to-end THz networks \cite{thz}.
THz communications could play a significant role in enabling the proliferation of UAS and HAPS, which are capable to guarantee lower latencies w.r.t. other NTN platforms operating at higher altitudes and provide  LoS air-to-ground paths. However, optimized positioning of such NTN platforms and proper beam management techniques are still required to provide reliable LoS links to ground users.
Furthermore, while the use of the THz band has been studied in the literature for inter-satellite communications, its feasibility for space-to-Earth links needs further investigation. The lack of compact and efficient THz sources and amplifiers remains the main challenge. Ultra-massive MIMO (UM-MIMO) techniques and reconfigurable electronic surfaces acting as smart reflectors could be effectively exploited. 

\textbf{\textit{Optical Wireless Communication (OWC)}} is a further 6G key technology candidate for being used in future NTN to achieve aggregate terabits-per-second capacity \cite{kau} and greater energy efficiency compared to RF-based solutions. OWC, including Free Space Optics (FSO) and Visible Light Communications (VLC), is used for inter-satellite communications. Massive R\&D activities are ongoing aimed at connecting small satellites and improving space-to-ground communications. On one hand, exploiting OWC for future 6G satellite platforms requires studies to overcome limitations due (i) to the interference of the solar background radiation, especially at low LEO orbits, and (ii) to difficulties in acquiring and maintaining strong and interference-free links in the presence of dense small satellite networks constellations with rapidly and frequently changing topology and line-of-sight loss due to mobility. High gain, multi-band, multi-functional, and multi-beam, smart steerable antennas can be viable solutions, as well as orbital path planning and ISL design joint optimization \cite{has}. On the other hand, the use of OWC for enabling space-to-ground links shows weaknesses that are inherent to the presence of atmospheric turbulence and adverse weather conditions in the lowest levels of the atmosphere. Therefore, efforts are still required to enable a transition from RF-based to non-RF-based solutions in the future 6G NTN. However, while waiting to define adequate modulation and error correction schemes or compensation of signal distortions to contrast the problems introduced by propagation through the atmosphere, hybrid solutions shall be considered, i.e., the literature proposes either the use of NR transmissions up to aerial platforms and then relaunch the signal in OWC mode towards satellites or leveraging communication modules that switch from NR to OWC and vice versa based on current atmospheric conditions.

\textbf{\textit{Intelligent Reflecting Surfaces (IRS),}} 
{also known as Reconfigurable Intelligent Surfaces (RIS), composed of a large array of individually configurable scattering elements, may actively control the signal propagation properties in favor of signal reception \cite{nyato2}. These can be employed in NTNs to improve the communication performance in a passive and energy-efficient way while meeting their typical size/weight/energy constraints. 
Interesting studies are proposing to jointly use RIS and THz band in LEO ISLs, to exploit IRS to enhance the security of satellite downlink communications via secure cooperative jamming, or to optimize IRS operation with predictive mobility compensation to increase SNR in IRS-assisted LEO satellite communications. 
Future studies are still needed in this area to overcome the current limitations related to channel models for NTNs and to the effects of harsh conditions on the design and operation of RISs.}

\textbf{\textit{Cloudification}} optimizes resource exploitation, saves operational costs, and ensures network flexibility while dealing with a massive number of clients and servers that 6G Networks will have to handle \cite{cloud}.
{Multi-access Edge Computing (}MEC{)} is a key technology to reduce processing latency for NTNs. The advantages associated with it will only be enabled by designing appropriate solutions for MEC services provisioning to NTN UEs and tasks processing directly at the NTNs.
A Cloud-native vision introduces new models for space business that improve the competitiveness of NTNs. In this view, virtualization techniques may provide satellites/nanosatellites with Digital Twins (DTs) hosted and properly orchestrated within a federated multi-tenant Ground Station Network (GSN) where GSs are edge nodes of a global terrestrial-satellite network. This allows to deliver 6G complex services by orchestrating and composing microservices associated with the different satellite DTs.

\textbf {\textit{Blockchain}} 
{can guarantee data security of 6G AI applications, improve network intelligence in big data processing, and protect data. 
Several ongoing studies highlight the opportunities offered by Blockchain to secure and prioritize multi-sensor satellite collaborative data exchanges, to introduce reputation systems when relaying data across multiple large constellations of small satellites and multiple TN-NTN integrated network segments, to verify access to LEO satellites under high location variability, to secure satellite swarms communications, and even turn space assets into secure digital tokens \cite{tork}. 
However, much remains to be investigated to deploy secure 6G NTN systems, wherein user privacy, entity authentication, and data protection/confidentiality/integrity are guaranteed via the Blockchain by NTN platforms regardless of their altitude and payload configuration.}

\textbf{\textit{Network and service orchestration and management}} 
{support multiple stakeholders and ensure network flexibility. In a 6G context, AI and machine learning are exploited to orchestrate and manage a fully intelligent network.
Network and service orchestration and management may also enable new use cases from the space by exploiting technologies for virtualization of network functions and device resources. 
In this view, future researches shall focus on techniques for suitably coping with the typical propagation delays of NTNs, especially in the presence of services requiring low latency, and on smart slicing solutions in stratified TN and NTN segments, that separate data flows of a slice {into} more micro-services to be delivered by satellites at different altitudes.}

\begin{figure}[h]
\centering
\includegraphics[scale=0.48]{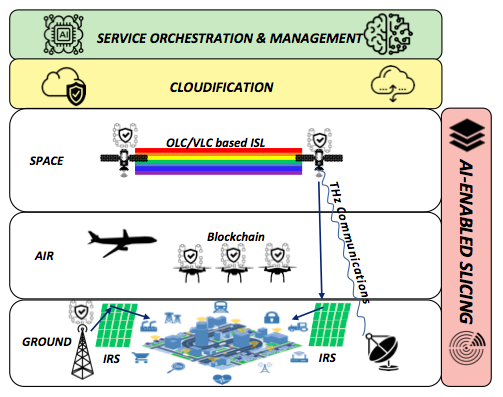}
\caption{{6G enabling technologies over space-air-ground networks.}}
\label{fig:6G}
\end{figure}

Figure \ref{fig:6G} depicts 6G enabling technologies and paradigms over space-air-ground networks.

\section{Conclusions}
\label{sec:8}

In this article, we surveyed main NR system design features covering the key innovative aspects introduced in the NR-enabled NTN system architecture and protocol stack to support 6G applications.
Moreover, we investigated the impact of NTN on NR procedures by highlighting adjustments, which have been suggested by 3GPP to enable NR in NTNs. Finally, we provided insights on the key enabling technologies for future 6G-oriented NTNs.

\renewenvironment{IEEEbiography}[1]
  {\IEEEbiographynophoto{#1}}
  {\endIEEEbiographynophoto}

\begin{IEEEbiography}
{Giuseppe Araniti} (araniti@unirc.it) received the Ph.D. degree in electronic engineering (2004) from University Mediterranea of Reggio Calabria, Italy, where he is Associate Professor of telecommunications. His major area of research is on 5G/6G networks including personal communications, enhanced wireless and satellite systems, traffic and radio resource management, D2D and M2M/MTC.
\\
\\
\textbf{Antonio Iera} (antonio.iera@dimes.unical.it) a Ph.D. degree from the University of Calabria in 1996. 
He currently holds the position of full professor of Telecommunications at the University of Calabria, Italy. His research interests include next generation mobile and wireless systems, and the Internet of Things.
\\
\\
\textbf{Sara Pizzi} (sara.pizzi@unirc.it) is an assistant professor in Telecommunications at University Mediterranea of Reggio Calabria, Italy, where she received the Ph.D. degree (2009) in Computer, Biomedical and Telecommunication Engineering. Her current research interests focus on NTN, RRM for multicast service delivery, D2D and MTC over 5G/B5G networks.
\\
\\
\textbf{Federica Rinaldi} (federica.rinaldi@unirc.it) is a Post-Doc Researcher at University Mediterranea of Reggio Calabria, Italy, where she received the Ph.D. in Information Engineering in 2021. Her current research interests include NTN, integrated terrestrial-NTN systems, RRM for multicast/broadcast service delivery, and D2D in 5G/B5G networks.

\end{IEEEbiography}

\end{document}